\documentclass{basi}

\newcommand{\msun}{\ensuremath{{{M}}_{\scriptscriptstyle \odot}}}

\newcommand{\beq}{\begin{equation}}
\newcommand{\eeq}{\end{equation}}
\def\Omm{{\Omega_m}}
\def\Ommz{{\Omega_m^{\,z}}}

\def\Omk{{\Omega_k}}
\def\Oml{{\Omega_{\Lambda}}}

\usepackage{txfonts}
\usepackage{graphicx}

\begin{document}

\title[Massive black hole seeds]{The formation and evolution of massive black hole seeds 
       in the early Universe}
\author[Priyamvada Natarajan]%
       {Priyamvada Natarajan,$^{1,2,3}$\thanks{e-mail:priyamvada.natarajan@yale.edu}\\
       $^1$Department of Astronomy, Yale University, 260 Whitney Avenue, New Haven, CT 06511\\
       $^1$Department of Physics, Yale University, P.O. Box New Haven, CT 06520\\
       $^3$Institute for Theory and Computation, Harvard University, 60 Garden Street, Cambridge MA 02138}

\pubyear{2011}
\volume{00}
\pagerange{\pageref{firstpage}--\pageref{lastpage}}
\date{Received 2011 February 27; accepted 2011 March 24}

\maketitle
\label{firstpage}

\begin{abstract}Tracking the evolution of high redshift seed black hole masses
to late times, we examine the observable signatures today.  These 
massive initial black hole seeds form at extremely high redshifts from the
direct collapse of pre-galactic gas discs. Populating dark matter
halos with seeds formed in this fashion, we follow the mass assembly history 
of these black holes to the present time using a Monte-Carlo merger
tree approach. Utilizing this formalism, we predict the black hole mass 
function at high redshifts and at the present time; the integrated mass density of
black holes in the Universe; the luminosity function of accreting black holes as a
function of redshift and the scatter in observed, local $M_{\rm bh}-\sigma$ relation. 
Comparing the predictions of the `light' seed model with these massive
seeds we find that significant differences appear predominantly at the low mass 
end of the present day black hole mass function. However, all our models predict 
that low surface brightness, bulge-less  galaxies with large discs are least likely to 
be sites for the formation of massive seed black holes at high redshifts. The efficiency 
of seed formation at high redshifts has a direct influence on the black hole
occupation fraction in galaxies at $z=0$. This effect is more
pronounced for low mass galaxies.  This is the key discriminant
between the models studied here and the Population III remnant `light' seed
model.  We find that there exists a population of low mass galaxies
that do not host nuclear black holes. Our prediction of the shape of
the $M_{\rm bh} - \sigma$ relation at the low mass end and increased
scatter has recently been corroborated by observations.
\end{abstract}

\begin{keywords}
 black holes -- galaxies: evolution -- galaxies: high redshift
\end{keywords}


\section{Introduction}\label{s:intro}

Demography of local galaxies suggests that most galaxies
harbour quiescent super-massive black holes (SMBHs) in their nuclei 
at the present time and that the mass of the hosted SMBH is correlated with 
properties of the host bulge. In fact, observational evidence points to the 
existence of a strong correlation between the mass of the central SMBH 
and the velocity dispersion of the host spheroid (Tremaine et al. 2002;
Ferrarese \& Merritt 2000, Gebhardt et al. 2003; Marconi \& Hunt 2003;
H\"aring \& Rix 2004; G\"ultekin et al. 2009) and possibly the host
halo (Ferrarese 2002) in nearby galaxies. These correlations are strongly 
suggestive of co-eval growth of the SMBH and the stellar component, likely via 
regulation of the gas supply in galactic nuclei from the earliest times 
(Haehnelt, Natarajan, Rees 1998; Silk \& Rees 1999; Kauffmann \& Haehnelt 2000; 
Fabian 2002; King 2003; Thompson, Quataert \& Murray 2005; 
Natarajan \& Treister 2009).

\section{Links between massive SMBH seeds, halo mass and spin}

Optically bright quasars powered by accretion onto black holes are now detected out 
to redshifts of $z > 6$ when the Universe was barely 7\% of its current age (Fan et al. 2004; 
2006).  The luminosities of these high redshift quasars imply black hole masses $M_{\rm
  BH} > 10^9\,M_{\odot}$. Models that describe the growth and
accretion history of supermassive black holes typically use as initial seeds the remnants
derived from Pop-III stars (e.g. Haiman \& Loeb 1998; Haehnelt, Natarajan \& Rees 1998).  
Assembling these large black hole masses by this early epoch starting from remnants of the first 
generation of metal free stars has been a challenge for models.  
Some suggestions to accomplish rapid growth invoke super-Eddington accretion rates for
brief periods of time (Volonteri \& Rees 2005). Alternatively, it has been
suggested that the formation of more massive seeds ab-initio through direct  
collapse of self-gravitating pre-galactic disks might offer a new channel as proposed by 
Lodato \& Natarajan 2006 [LN06]. This scenario alleviates the problem of building up
supermassive black hole masses to the required values by $z = 6$.

We focus on the main features of massive seed models in this review.  Most aspects of the 
evolution and assembly history of this scenario have been explored in detail in 
Volonteri \& Natarajan (2009) and Volonteri, Lodato \& Natarajan (2008). In these models, at 
early times the properties of the  assembling SMBH seeds are more tightly coupled to properties 
of the dark matter halo as their growth is driven by the merger history of halos. However, at later times, when 
the merger rates are low, the final mass of the SMBH is likely to be more tightly coupled to the small scale 
local baryonic distribution.  The relevant host dark matter halo property at high redshifts in this 
picture is the spin.

In a physically motivated model for the 
formation of heavy SMBH seeds (in contrast to the lower mass remnant
seeds from Population III stars) as described in LN06, there is a limited
range of halo spins and halo masses that are viable sites for the
formation of seeds. In this picture, massive seeds with $M\approx 10^5-10^6M_{\odot}$ 
can form at high redshift ($z>15$), when the
intergalactic medium has not been significantly enriched by metals
(Koushiappas, Bullock \& Dekel 2004; Begelman, Volonteri \& Rees 2006;  LN06; 
Lodato \& Natarajan 2007).  As derived in LN06, the
development of non-axisymmetric spiral structures drives mass infall
and accumulation in a pre-galactic disc with primordial
composition. The mass accumulated in the center of the halo (which
provides an upper limit to the SMBH seed mass) is given by:
\begin{equation}
M_{\rm BH}= m_{\rm d}M_{\rm halo}\left[1-\sqrt{\frac{8\lambda}{m_{\rm d}Q_{\rm c}}\left(\frac{j_{\rm d}}{m_{\rm d}}\right)\left(\frac{T_{\rm gas}}{T_{\rm vir}}\right)^{1/2}}\right] 
\label{mbh}
\end{equation}
for 
\begin{equation}
\lambda<\lambda_{\rm max}=m_{\rm d}Q_{\rm c}/8(m_{\rm d}/j_{\rm d}) (T_{\rm
  vir}/T_{\rm gas})^{1/2}
\label{lambdamax} 
\end{equation}
and $M_{\rm BH}=0$ otherwise. Here $\lambda_{\rm max}$ is the maximum
halo spin parameter for which the disc is gravitationally unstable,
$m_d$ is the gas fraction that participates in the infall and $Q_{\rm
c}$ is the Toomre parameter.  The efficiency of SMBH formation is
strongly dependent on the Toomre parameter $Q_{\rm c}$, which sets the
frequency of formation, and consequently the number density of SMBH
seeds. The efficiency of the seed assembly process ceases at large halo
masses, where the disc undergoes fragmentation instead. This occurs
when the virial temperature exceeds a critical value $T_{\rm max}$,
given by:
\begin{equation}
\frac{T_{\rm max}}{T_{\rm gas}}=\left(\frac{4\alpha_{\rm c}}{m_{\rm
d}}\frac{1}{1+M_{\rm BH}/m_{\rm d}M_{\rm halo}}\right)^{2/3},
\label{frag}
\end{equation}
where $\alpha_{\rm c}\approx 0.06$ is a dimensionless parameter measuring the
critical gravitational torque above which the disc fragments.
The remaining relevant parameters are assumed to have typical values: $m_{\rm d}=j_{\rm d}=0.05$,
$\alpha_{\rm c}=0.06$ for the $Q_{\rm c}=2$ case.
The gas has a temperature $T_{\rm gas}=5000$K.

\begin{figure}
\centerline{\includegraphics[width=14cm,clip=]{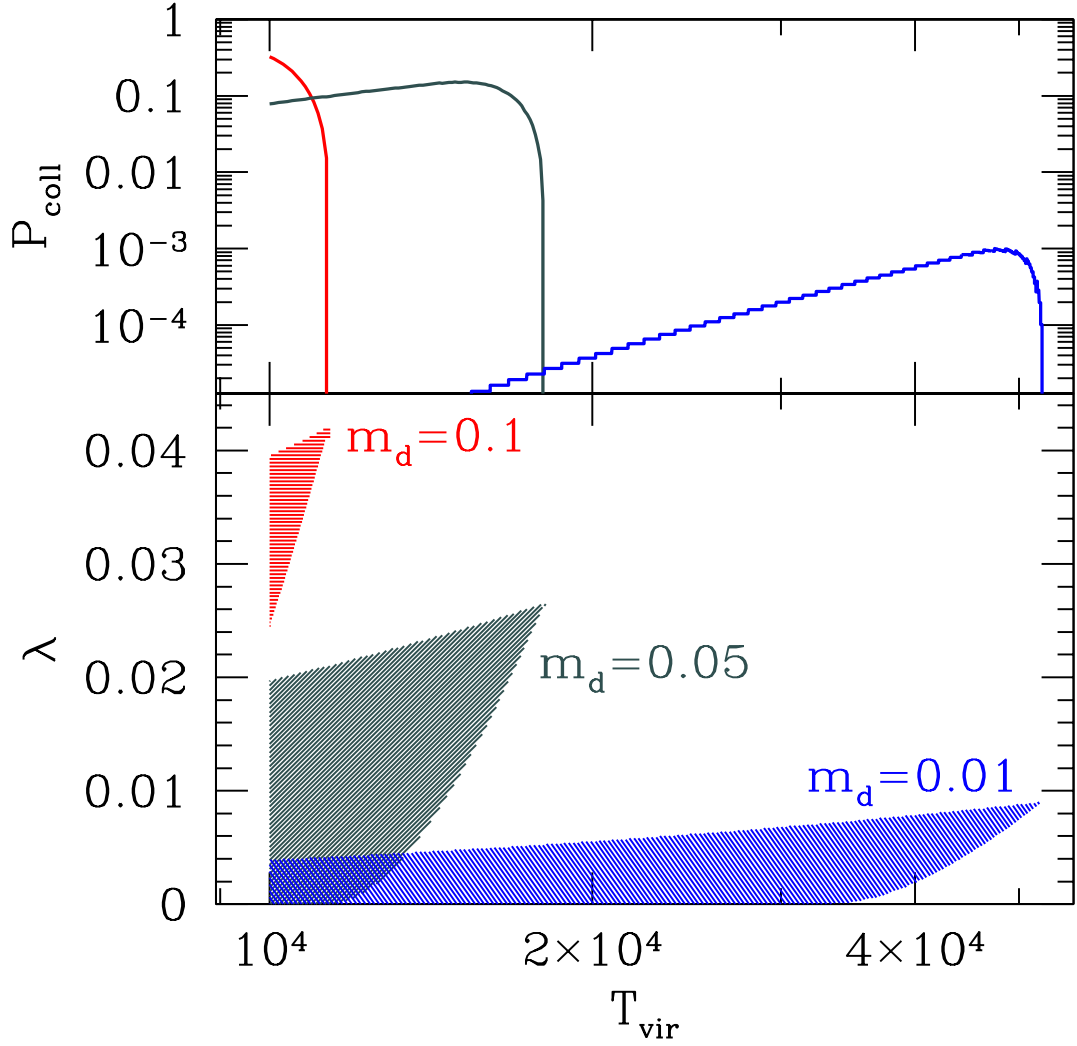}}
\caption{Parameter space (virial temperature, spin parameter) for
  SMBH formation. Halos with $T_{vir}>10^4$ K
  at $z=15$ are picked to participate in the infall ($m_{\rm d}$). 
  The shaded areas in the bottom panel show the range of virial temperatures and spin parameters 
  where discs are Toomre unstable and the joint conditions, $\lambda<\lambda_{\rm max}$
  (equation~\ref{lambdamax}) and $T_{\rm vir}<T_{\rm max}$ (equation~\ref{frag}, showing the minimum spin 
  parameter, $\lambda_{\rm min}$ value below which the disc is globally prone to fragmentation) are fulfilled. 
  The top panel shows the probability of SMBH formation and is obtained by integrating the lognormal distribution 
  of spin parameters between $\lambda_{\rm min}$ and  $\lambda_{\rm max}$.}
\label{Pcoll_lambda}
\end{figure}

To summarize, every dark matter halo is characterized by its mass $M$
(or virial temperature $T_{\rm vir}$) and by its spin parameter
$\lambda$.  If $\lambda<\lambda_{\rm max}$ (see equation~\ref{lambdamax}) and $T_{\rm
vir}<T_{\rm max}$ (equation~\ref{frag}), then a seed SMBH forms in the centre. 
Hence SMBHs form (i) only in halos within a given range of virial
temperatures, and hence, halo masses, and (ii) only within a narrow range of 
spin parameters, as shown in Figure~\ref{Pcoll_lambda}. High values of the spin
parameter, leading most likely to disk-dominated galaxies, are strongly disfavored as
seed formation sites in this model, and in models that rely on global dynamical 
instabilities (Volonteri \& Begelman 2010).

\section{The evolution of seed black holes}

We follow the evolution of the MBH population resulting from the seed
formation process delineated above in a $\Lambda$CDM Universe. Our
approach is similar to the one described in Volonteri, Haardt \&
Madau (2003). We simulate the merger history of present-day halos with
masses in the range $10^{11}<M<10^{15}\,\msun$ starting from $z=20$,
via a Monte Carlo algorithm based on the extended Press-Schechter
formalism. Every halo entering the merger tree is assigned a spin parameter 
drawn from the lognormal $P(\lambda)$ distribution of simulated LCDM halos. 
Recent work on the fate of halo spins during mergers in cosmological simulations has led to
conflicting results: Vitvitska et al.  (2002) suggest that the spin
parameter of a halo increases after a major merger, and the angular
momentum decreases after a long series of minor mergers; D'Onghia \&
Navarro (2007) find instead no significant correlation between spin
and merger history. Given the unsettled nature of this matter,  we simply assume that 
the spin parameter of a halo is not modified by its merger history.

When a halo enters the merger tree we assign seed MBHs by determining
if the halo meets all the requirements described in Section 2 for the
formation of a central mass concentration. As we do not
self-consistently trace the metal enrichment of the intergalactic
medium, we consider here a sharp transition threshold, and assume that
the MBH formation scenario suggested by Lodato \& Natarajan ceases at
$z\approx 15$ (see also Sesana 2007; Volonteri 2007). 
At $z>15$, therefore, whenever a new halo appears in the merger tree 
(because its mass is larger than the mass resolution), or a pre-existing 
halo modifies its mass by a merger, we evaluate if the gaseous component 
meets the conditions for efficient transport of angular momentum to create a
large inflow of gas which can either form a MBH seed, or feed one if
already present. 

\begin{figure}   
\centerline{\includegraphics[width=14cm,clip=]{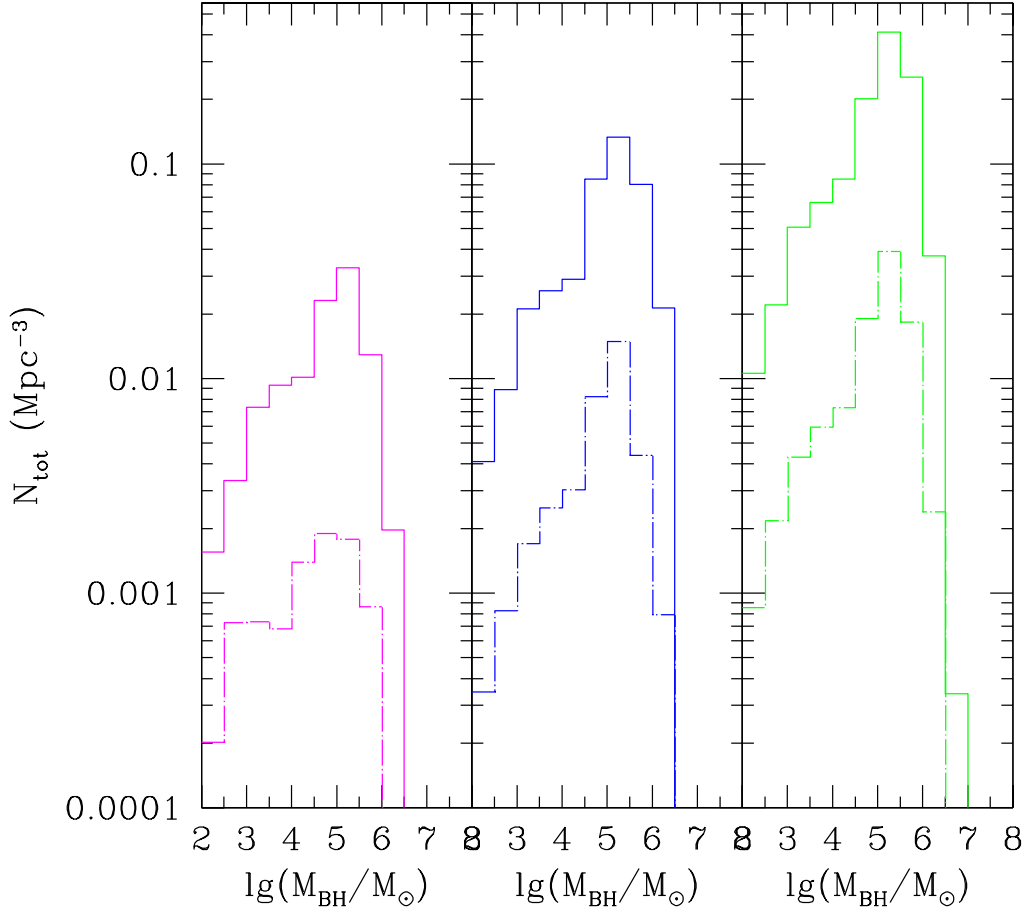}} 
   \caption{Mass function of MBH seeds in the three Q-models that
differ in seed formation efficiency. Left panel: $Q_{\rm c}=1.5$ (the
least efficient model A), middle panel: $Q_{\rm c}=2$ (intermediate
efficiency model B), right panel: $Q_{\rm c}=3$ (highly efficient
model C). Seeds form at $z>15$ and this channel ceases at $z =
15$. The solid histograms show the total mass function of seeds formed
by $z=15$, while the dashed histograms refer to seeds formed at a
specific redshift, $z=18$.}
   \label{fig2}
\end{figure}

The efficiency of MBH formation is strongly dependent on a critical value of 
the Toomre parameter $Q_{\rm c}$, which sets the frequency of formation, and
consequently the number density of MBH seeds. We investigate the
influence of this parameter in the determination of the global
evolution of the MBH population. Figure~\ref{fig2} shows the number
density of seeds formed in three different models with varying efficiency, 
with $Q_{\rm c}=1.5$ (low efficiency model A), $Q_{\rm c}=2$ (intermediate 
efficiency model B), and $Q_{\rm c}=3$ (high efficiency model C). The solid 
histograms show the total mass function of seeds formed by $z=15$ when this
 formation channel ceases, while the dashed histograms refer to seeds formed in a
specific redshift slice at $z=18$.  The number of seeds changes by
about one order of magnitude from the least efficient to the most
efficient model, consistent with the probabilities shown in Figure~1.

We assume that, after seed formation ceases, the $z<15$ population of
MBHs evolves according to a ``merger driven scenario", as described in Volonteri (2006). 
We assume that during major mergers MBHs accrete gas mass
that scales with the fifth power of the circular velocity
(or equivalently the velocity dispersion $\sigma_c$) of the host halo
(Ferrarese 2002). We thus set the final mass of the MBH at the end of the
accretion episode to 90\% of the mass predicted by the $M_{\rm BH}-\sigma_c$
correlation, assuming that the scaling does not evolve with
redshift. Major mergers are defined as mergers between two dark matter
halos with mass ratio between 1 and 10. BH mergers contribute to the
mass addition of the remaining 10\%.

We briefly outline the merger scenario calculation here. The merger rate of halos can be estimated 
using equation~1 of Fakhouri, Ma \& Boylan-Kolchin (2010), where a simple fitting formula is 
derived from large LCDM simulations.
The merger rate per unit redshift and mass ratio ($\xi$) at fixed halo mass is given by:
\beq
\frac{dN_m}{d\xi dz}(M_h) = A\left(\frac{M_h}{10^{12} M_0}\right)^{\alpha}\xi^{\beta}\exp\left[\left(\frac{\xi}{\tilde{\xi}}\right)^{\gamma}\right](1+z)^{\eta}.
\label{mjm}
\eeq
with A = 0.0104, $\alpha=0.133$, $\beta=-1.995$, $\gamma=0.263$, $\eta=0.0993$ and $\tilde{\xi}=9.72\times10^{-3}$. We
can integrate the merger rate between $z=0$ and say, $z=3$, for major mergers. This gives the number 
of major mergers a halo of a given mass experiences between $z=0$ and $z=3$. 
Halo mass can be translated into virial circular velocity:
\begin{equation}
 V_{\rm c}= 142 {\rm km/s} \left[\frac{M_{\rm h}}{10^{12} \ M_{\odot} }\right]^{1/3} 
\left[\frac {\Omm}{\Ommz}\ \frac{\Delta_{\rm c}} {18\pi^2}\right]^{1/6} 
(1+z)^{1/2},  
\end{equation}
where $\Delta_{\rm c}$ is the over--density at virialization relative 
to the critical density. 
For a WMAP5 cosmology we adopt here the  fitting
formula  $\Delta_{\rm c}=18\pi^2+82 d-39 d^2$ (Bryan \& Norman 1998), 
where $d\equiv \Ommz-1$ is evaluated at the collapse redshift, so
that $ \Ommz={\Omm (1+z)^3}/({\Omm (1+z)^3+\Oml+\Omk (1+z)^2})$.  It is well known that the 
major merger rate is an increasing function of halo mass or circular velocity. In fact we 
find that the expected number of mergers between $z=0$ and $z=3$ with mass ratio $\xi >0.3$ is
$\simeq$~0.4 for $M_h=10^8 \msun$, $\simeq$~0.5 for $M_h=10^9 \msun$, $\simeq$~0.7 for $M_h=10^{10} \msun$, $\simeq$~1.0 for $M_h=10^{11} \msun$,
$\simeq$~1.4 for $M_h=10^{12} \msun$, $\simeq$~1.8 for $M_h=10^{13} \msun$.  

In order to calculate the luminosity function of active black holes and
to follow the black hole mass growth during each accretion event, we
also need to calculate {the mass inflow rate.} This is assumed to scale with the Eddington rate for the
MBH, and is based on the results of merger simulations, which 
heuristically track accretion onto a central MBH (Di Matteo, Springel \& Hernquist 2005; 
Hopkins et al. 2005; Sijacki et al. 2007).  The time spent by a given simulated AGN 
at a given bolometric luminosity\footnote{We convert accretion rate 
into luminosity assuming that the radiative efficiency equals the 
binding energy per unit mass of a particle in the last stable circular 
orbit. We associate the location of the last stable circular orbit with 
the spin of the MBHs, by self-consistently tracking the evolution of 
black hole spins throughout our calculations (Volonteri 2006). We set 20\% as the maximum value of the 
radiative efficiency, corresponding to a spin slightly below the 
theoretical limit for thin disc accretion (Thorne 1974). } 
per logarithmic interval is approximated by Hopkins et al. (2005) as:
\begin{equation}
\label{eq:dtdlogL }
\frac{{d}t}{{d}L}=|\alpha|t_Q\, L^{-1}\, \left(\frac{L}{10^9L_\odot}\right)^\alpha,
\end{equation}
where $t_Q\simeq10^9$ yr, and $\alpha=-0.95+0.32\log(L_{\rm
peak}/10^{12} L_\odot)$. Here $L_{\rm peak}$ is the luminosity of the
AGN at the peak of its activity.  Hopkins et al. (2006) show that
approximating $L_{\rm peak}$ by the Eddington luminosity of the MBH
at its final mass (i.e., when it sits on the $M_{\rm BH}-\sigma_c$
relation) compared to computing the peak luminosity with equation~(6)
above gives the same result and in fact, the difference between these
two cases is negligible. Volonteri, Salvaterra \& Haardt (2006) derive the following
simple differential equation to express the instantaneous accretion
rate ($f_{\rm Edd}$,in units of the Eddington rate) for a MBH of mass
$M_{\rm BH}$ in a galaxy with velocity dispersion $\sigma_c$:
\begin{equation}
\label{eq:doteddratio }
\frac{{d}f_{\rm Edd}(t)}{{d}t}=\frac{ f_{\rm Edd}^{1-\alpha}(t) }{|\alpha|
  t_Q} \left(\frac{\epsilon \dot{M}_{\rm Edd} c^2}{10^9L_\odot}\right)^{-\alpha},
\end{equation}
where here $t$ is the time elapsed from the beginning of the accretion event.
Solving this equation provides us with the instantaneous Eddington ratio for a
given MBH at a specific time, and therefore we can self-consistently follow
the MBH mass. We set the Eddington ratio $f_{\rm Edd}=10^{-3}$ at
$t=0$. This same type of accretion is assumed to occur, at $z>15$,
following a major merger in which a MBH is not fed by disc
instabilities. 

\begin{figure}   
\centerline{\includegraphics[width=11.3cm,clip=]{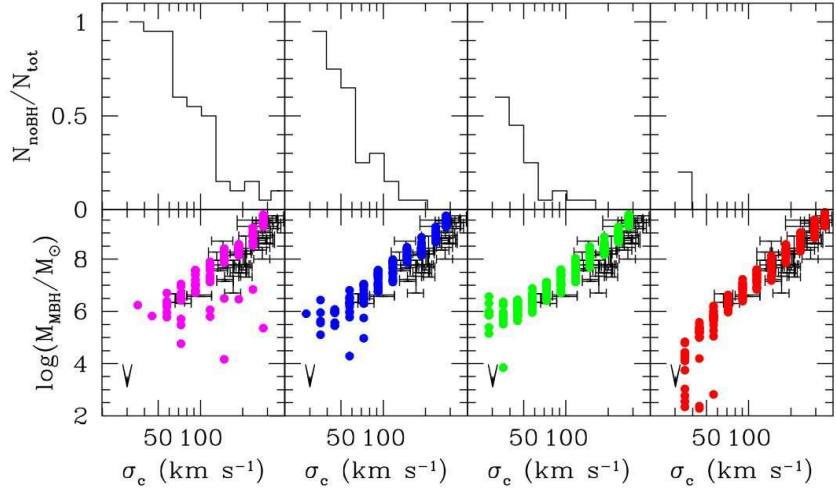}} 
\caption{ The $M_{\rm bh}-$velocity dispersion ($\sigma_c$)
     relation at $z=0$. Every circle represents the central MBH in a
     halo of given $\sigma_c$.  Observational data are marked by their
     quoted errorbars, both in $\sigma_c$, and in $M_{\rm bh}$
     (Tremaine et al. 2002).  Left to right panels: $Q_{\rm c}=1.5$,
     $Q_{\rm c}=2$, $Q_{\rm c}=3$, Population III star seeds.  {\it
     Top panels:} fraction of galaxies at a given velocity dispersion
     which {\bf do not} host a central MBH.}
\label{fig3}
\end{figure}

\section{Results}

The repercussions of different initial efficiencies for seed formation
for the overall evolution of the MBH population stretch from
high-redshift to the local Universe.
Detection of gravitational waves from seeds merging at the redshift of
formation (Sesana 2007) is probably one of the best ways to
discriminate among formation mechanisms. On the other hand, the
imprint of different formation scenarios can also be sought in
observations at lower redshifts. The various seed formation scenarios
have distinct consequences for the properties of the MBH population at
$z=0$. 

\subsection{Low redshift predictions}

\subsubsection{Supermassive black holes in dwarf galaxies}

Obviously, a higher density of
MBH seeds implies a more numerous population of MBHs at later times,
which can produce observational signatures in statistical
samples. More subtly, the formation of seeds in a $\Lambda$CDM
scenario follows the cosmological bias. As a consequence, the
progenitors of massive galaxies (or clusters of galaxies) have a
higher probability of hosting MBH seeds (cf. Madau \& Rees 2001). In
the case of low-bias systems, such as isolated dwarf galaxies, very
few of the high-$z$ progenitors have the deep potential wells needed
for gas retention and cooling, a prerequisite for MBH formation. 
In the lowest efficiency model A, for example, a galaxy needs of order 25 massive progenitors (mass
above $\sim10^7\msun$) to ensure a high probability of seeding within
the merger tree. In model C, instead, the requirement drops to 4
massive progenitors, increasing the probability of MBH formation in
lower bias halos.

The signature of the efficiency of the formation of MBH seeds will
consequently be stronger in isolated dwarf galaxies. Figure \ref{fig3}
(bottom panel) shows a comparison between the observed $M_{\rm
BH}-\sigma$ relation and the one predicted by our models (shown with
circles), and in particular, from left to right, the three models
based on the LN06 and Lodato \& Natarajan (2007) seed masses with $Q_{\rm c}=1.5$, 2 and
3, and a fourth model based on lower-mass Population III star
seeds. The upper panel of Figure \ref{fig3} shows the fraction
of galaxies that {\bf do not} host any massive black holes for different
velocity dispersion bins. This shows that the fraction of galaxies
without a MBH increases with decreasing halo masses at $z = 0$. 
A larger fraction of low mass halos are devoid of central black holes for
lower seed formation efficiencies. Note that this is one of the key
discriminants between our models and those seeded with Population III
remnants. As shown in Figure~3, there are practically no galaxies without
central BHs for the Population III seeds.

We can therefore make quantitative predictions for the local occupation fraction of MBHs. 
Our model A predicts that below $\sigma_c\approx 60\,{\rm kms}^{-1}$ the
probability of a galaxy hosting a MBH is negligible. With increasing
MBH formation efficiencies, the minimum mass for a galaxy that hosts a
MBH decreases, and it drops below our simulation limits for model
C. On the other hand, models based on lower mass Population III star
remnant seeds, predict that massive black holes might be present even
in low mass galaxies. Our predictions have been corroborated by recent
observations of low mass galaxies (Kormendy \& Bender 2011).

Although there are degeneracies in our modeling (e.g., between the
minimum redshift for BH formation and the instability
criterion), the BH occupation fraction and the masses of the BHs in
dwarf galaxies are the key diagnostics.  An additional caveat worth mentioning is the
possibility that a galaxy is devoid of a central MBH because of
dynamical ejections (due to either the gravitational recoil or
three-body scattering). The signatures of such dynamical interactions
should be more prominent in dwarf galaxies, but ejected MBHs would
leave observational signatures on their hosts (G{\"u}ltekin et al. in
prep.). On top of that, Schnittman (2007) and Volonteri, Lodato \& Natarajan (2008)
agree in considering the recoil a minor correction to the overall
distribution of the MBH population at low redshift (cf. figure 4 in
Volonteri 2007).

Additionally, as MBH seed formation requires halos with low angular
momentum (small spin parameter), we envisage that low surface
brightness, bulge-less galaxies with large spin parameters (i.e. large
discs) are systems where MBH seed formation is less
probable. 
Furthermore, bulgeless galaxies are believed to have preferentially 
quieter merger histories and are unlikely to have experienced 
major mergers that could have brought in a MBH from a companion
galaxy. 

 \begin{figure}   
\centerline{\includegraphics[width=14 cm]{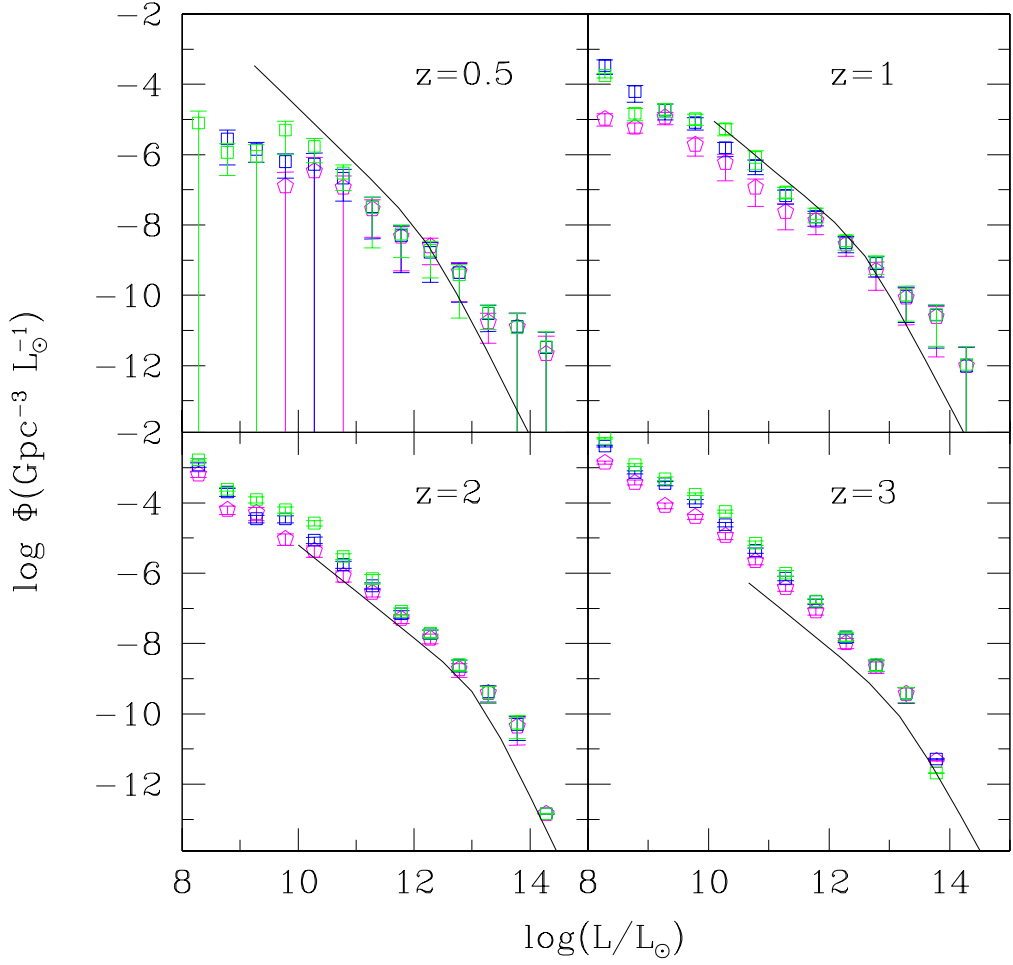}} 
   \caption{Predicted bolometric luminosity functions at different
redshifts with observational data over-plotted. All 3 models match the observed bright end of the LF at
high redshifts and predict a steep slope at the faint end down to $z =
1$. The 3 models are not really distinguishable with the LF. However
at low redshifts, for instance at $z = 0.5$, all 3 models are
significantly flatter at both high and low luminosities and do not
adequately match the current data. As discussed in the text, the LF is
strongly determined by the accretion prescription, and what we see here
is simply a reflection of that fact.}
   \label{fig4}
\end{figure}

\subsubsection{Comoving mass density of black holes}
 
\begin{figure}   
\centerline{\includegraphics[width=14cm]{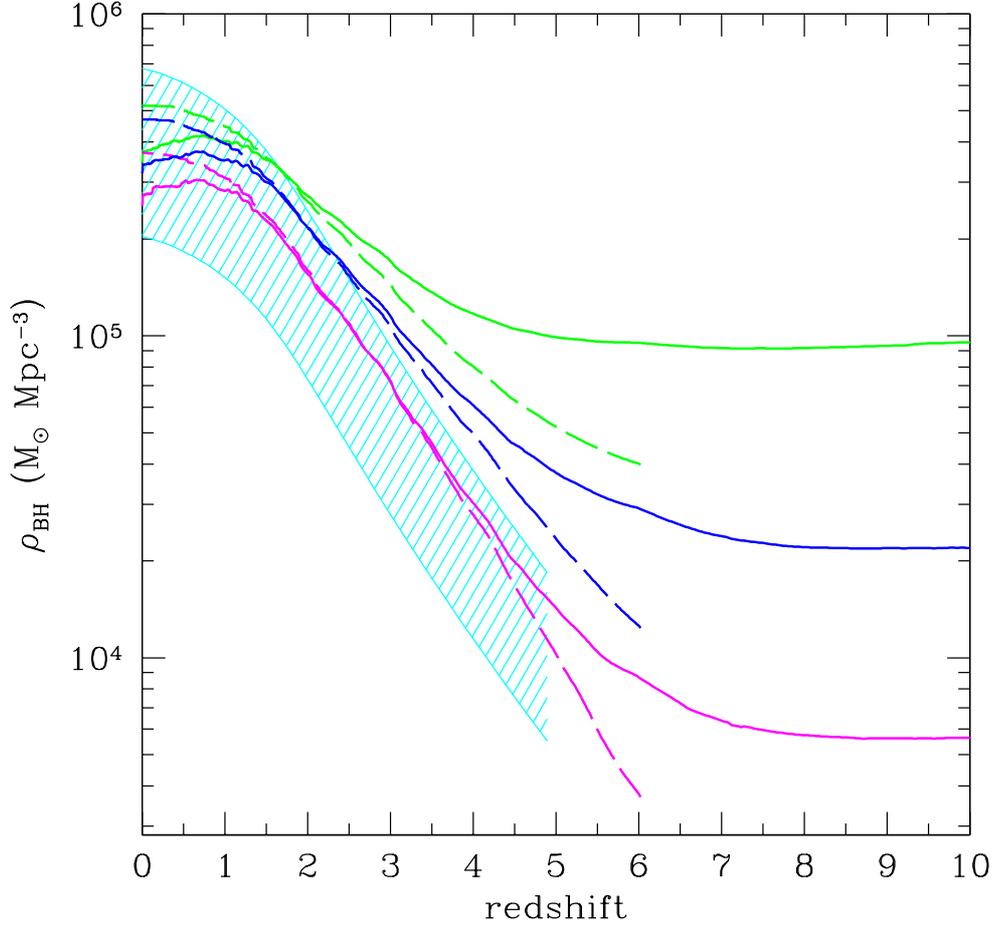}} 
   \caption{Integrated black hole mass density as a function of
     redshift. Solid lines: total mass density locked into nuclear
     black holes.  Dashed lines: integrated mass density {\rm
     accreted} by black holes.  Models based on BH remnants of
     Population III stars (lowest curve), $Q_{\rm c}=1.5$ (middle
     curve) and $Q_{\rm c}=2$ (upper curve).  Shaded area: constraints from So{\l}tan-type
     arguments, where we have varied the radiative efficiency from a
     lower limit of 6\% (applicable to Schwarzschild MBHs, upper
     envelope of the shaded area), to about 20\%. All 3 massive seed formation models are in comfortable
     agreement with the mass density obtained from integrating the
     optical luminosity functions of quasars.}
   \label{fig5}
\end{figure}

Since during the quasar epoch MBHs increase their mass by a large factor,
signatures of the seed formation mechanisms are likely more evident at
{\it earlier epochs}. We compare in Figure~\ref{fig5} the integrated
comoving mass density in MBHs to the expectations from So{\l}tan-type
arguments, assuming that quasars
are powered by radiatively efficient flows (for details, see Yu \& Tremaine 2002; 
Elvis, Risaliti \&  Zamorini 2002; Marconi et al. 2004). While during and after the quasar
epoch the mass densities in models A, B, and C differ by less than a
factor of 2, at $z>3$ the differences are more pronounced.

A very efficient seed MBH formation scenario can lead to a very large
BH density at high redshifts. For instance, in the highest efficiency
model C with $Q_{\rm c}=3$, the integrated MBH density at $z=10$ is
already $\sim 25\%$ of the density at $z=0$. The plateau at $z>6$ is
due to our choice of scaling the accreted mass with the $z=0$ $M_{\rm
bh}-\sigma$ relation. Since in our models we let MBHs
accrete mass that scales with the fifth power of the circular
velocity of the halo, the accreted mass is a small fraction of the MBH
mass (see the discussion in (Marulli et al. 2006), and the overall
growth remains small, as long as the mass of the seed is larger than
the accreted mass, which, for our assumed scaling, happens whenever  the
mass of the halo is below a few times $10^{10}\msun$. The comoving
mass density, an integral constraint, is reasonably well determined
out to $z = 3$ but is poorly known at higher redshifts. All models
appear to be satisfactory and consistent with current observational
limits (shown as the shaded area).

\subsubsection{Black hole mass function at $z=0$}

\begin{figure}   
\centerline{\includegraphics[width=14cm]{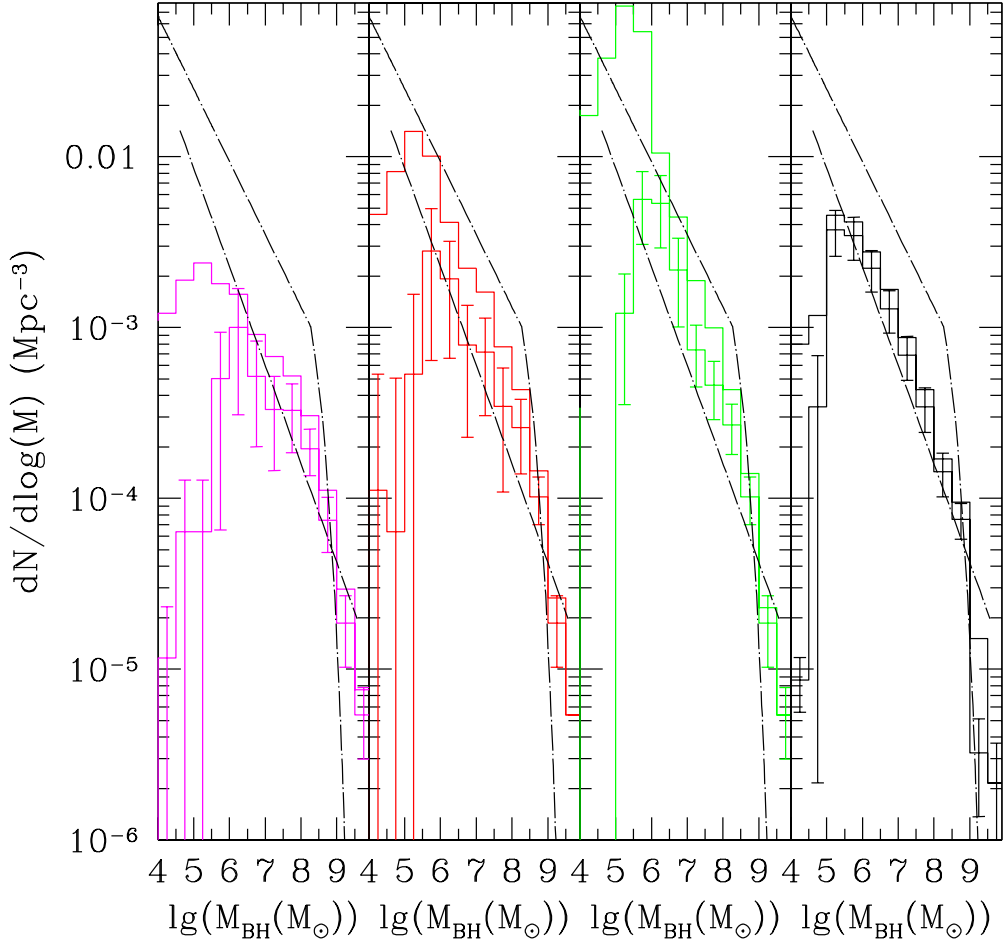}} 
   \caption{Mass function of black holes at z=0. Histograms represent
     the results of our models, including central galaxies only (lower
     histograms with error bars), or including satellites in groups and
     clusters (upper histograms). Left panel: $Q_{\rm c}=1.5$, mid-left
     panel: $Q_{\rm c}=2$, mid-right panel: $Q_{\rm c}=3$, right panel:
     models based on BH remnants of Population III stars.  
     Upper dashed line: mass function derived from combining the velocity
     dispersion function of Sloan galaxies (Sheth et al. 2003, where
     we have included the late-type galaxies extrapolation), and BH
     mass-velocity dispersion correlation (e.g., Tremaine et
     al. 2002). Lower dashed line: mass function derived using the
     Press-Schechter formalism from Jenkins et al. (2001) in
     conjunction with the $M_{\rm BH} -\sigma$ relation (Ferrarese
     2002).}
   \label{fig6}
\end{figure}

One of the key diagnostics is the comparison of the measured and
predicted BH mass function at $z = 0$ for our 3 models. In
Figure~\ref{fig6}, we show (from left to right, respectively) the mass
function predicted by models A, B, C and Population III remnant seeds
compared to that obtained from measurements.  The histograms show the
mass function obtained with our models (where the upper histogram
includes all the black holes while the lower one only includes black
holes found in central galaxies of halos in the merger-tree
approach). The two lines are two different estimates of the observed
black hole mass function. In the upper one, the measured velocity
dispersion function for nearby late and early-type galaxies from the
SDSS survey (Bernardi et al. 2003; Sheth et al. 2003) has been convolved with the
measured $M_{\rm BH} - \sigma$ relation. We note here that the scatter
in the $M_{\rm bh} - \sigma$ relation is not explicitly included in
this treatment, however the inclusion of the scatter is likely to
preferentially affect the high mass end of the BHMF, which provides
stronger constraints on the accretion histories than do the seed
masses. It has been argued by Tundo et al. (2007), Bernardi et al. (2007) 
and Lauer et al. (2007) 
that the BH mass function differs if the bulge mass is used instead of the 
velocity dispersion in relating the BH mass to the host galaxy. 
Since our models do not trace the formation and growth of stellar bulges 
in detail, we are restricted to using the velocity dispersion in our analysis.

The lower dashed curve is an alternate theoretical estimate of the BH mass 
function derived using the Press-Schechter formalism from  Jenkins et al. (2001) 
in conjunction with the observed $M_{\rm BH} -\sigma$ relation. Selecting only the
central galaxies of halos in the merger-tree approach adopted here
(lower histograms) is shown to be equivalent to this analytical
estimate, and this is clearly borne out in the plot.
When we include black holes in satellite galaxies (upper histograms,
cf. the discussion in Volonteri, Haardt \& Madau 2003) the predicted
mass function moves towards the estimate based on SDSS galaxies. The
higher efficiency models clearly produce more BHs. At higher
redshifts, for instance at $z = 6$, the mass functions of active MBHs
predicted by all models are in very good agreement, in particular for
BH masses larger than $10^6\,\msun$, as it is the growth by accretion
that dominates the evolution of the population. At the highest mass
end ($>10^9\,\msun$) model A lags behind models B and C, although we
stress once again that our assumptions for the accretion process are
very conservative.

The {\it relative} differences between models A, B, and C at the
low-mass end of the mass function, however, are genuinely related to
the MBH seeding mechanism (see also Figures~\ref{fig3} and
~\ref{fig5}). In model A, simply, fewer galaxies host a MBH, hence
reducing the overall number density of black holes. Although our
simplified treatment does not allow robust quantitative predictions,
the presence of a ``bump'' at $z = 0$ in the MBH mass function at the
characteristic mass that marks the peak of the seed mass function
(cf. Figure~\ref{fig2}) is a sign of highly efficient formation of
massive seeds (i.e., much larger mass than, for instance,
Population III remnants). The higher the efficiency of seed formation,
the more pronounced is the bump (note that the bump is most prominent
for model C). Since current measurements of MBH masses extend barely
down to $M_{\rm bh}\sim 10^6 \msun$, this feature cannot be
observationally tested with present data, but future campaigns, with
the Giant Magellan Telescope or JWST, are likely to extend the mass
function measurements to much lower black hole masses.

\subsection{Predictions at high redshift}

\subsubsection{The luminosity function of accreting black holes}

Turning to the global properties of the MBH population, as suggested
by Yu \& Tremaine (2002), the mass growth of the MBH population at
$z<3$ is dominated by the mass accreted during the bright epoch of
quasars, thus washing out most of the imprint of initial conditions.
This is evident when we compute the luminosity function of AGN.
Clearly the detailed shape of the predicted luminosity function
depends most strongly on the accretion prescription used. With our
assumption that the gas mass accreted during each merger episode is
proportional to $V_c^5$, we find that distinguishing between the
various seed models is difficult. As shown in Figure~\ref{fig4}, all 3
models reproduce the bright end of the observed bolometric LF
(Hopkins, Richards \& Hernquist 2007) at higher redshifts (marked as the solid curve in
all the panels), and predict a fairly steep faint end that is as yet
undetected. All models fare less well at low redshift, shown in
particular at $z = 0.5$. This could be due to the fact that we have
used a single accretion prescription to model growth at all times.
On the other hand, the decline in the available gas supply at
low redshifts (since the bulk of the gas has been consumed before this
epoch by star formation activity) likely changes the radiative
efficiency of these systems. Besides, observations suggest a sharp
decline in the number of actively accreting black holes at low
redshifts at different wave-lengths, produced most probably by changes
in the accretion flow as a result of changes in the geometry of the
nuclear regions of galaxies. In fact, all 3 of our models
under-predict the slope at the faint end. There are three other effects
that could cause this flattening of the LF at the faint end at low
redshift for our models: (i) not having taken into account the result of
on-going mergers and the fate of satellite galaxies; (ii) the number
of realizations generated and tracked is insufficient for statistics,
as evidenced by the systematically larger errorbars and (iii) more
importantly, it is unclear if merger-driven accretion is indeed the
trigger of BH fueling in the low redshift Universe. We note that the 3
massive seed models and Population III seed model cannot be
discriminated by the LF at high redshifts. Models B and C are also in
agreement viz-a-viz the predicted BH mass function at $z = 6$ (see Figure~2), even
assuming a very high radiative efficiency (up to 20\%), while model A might
need less severe assumptions, in particular for BH masses larger than
$10^7\,\msun$.

\section{Conclusions}

Ih this review, we outline massive black hole seed formation models 
and focus on the predictions made by these at high and low redshift.
While the errors on mass determinations of local black holes
are large at the present time, definite trends with host galaxy
properties are observed. The tightest correlation appears to be between
the BH mass and the velocity dispersion of the host spheroid. Starting
with the ab-initio black hole seed mass function computed in the
context of direct formation of central objects from the collapse of
pre-galactic discs in high redshift halos, we follow the assembly
history to late times using a Monte Carlo merger tree approach. Key to
our calculation of the evolution and build-up of mass is the
prescription that we adopt for determining the precise mass gain
during a merger. Motivated by the phenomenological observation of
$M_{\rm BH} \propto V_{\rm c}^5$, we assume that this proportionality
carries over to the gas mass accreted in each step. With these
prescriptions, a range of predictions can be made for the mass
function of black holes at high and low $z$, and for the integrated mass
density of black holes, all of which are observationally
determined. We evolve 3 models, designated model A, B and C, which
correspond to increasing efficiencies respectively for the formation
of seeds at high redshift. These models are compared to one in which
the seeds are remnants of Population III stars. 

It is important to
note here that one major uncertainty prevents us from making more
concrete predictions: the unknown metal enrichment history of
the Universe. Key to the implementation of our models is the choice of
redshift at which massive seed formation is quenched. The direct
seed formation channel described here ceases to operate once the Universe
has been enriched by metals that have been synthesized by the 
first generation of stars. Once metals are available
in the Inter-Galactic Medium, gas cooling is much more efficient and hydrogen
in either atomic or molecular form is no longer the key player. In this 
work, we have assumed this transition redshift to be $z = 15$. The efficiency 
of MBH formation and the transition redshift are somehow degenerate (e.g., 
a model with $Q=1.5$ and enrichment redshift $z=12$ is halfway between 
model A and model B); if other constraints on this redshift were available we 
could considerably tighten our predictions. 
 
Below we list our predictions and compare how they fare with respect
to current observations. The models investigated here clearly differ
in predictions at the low mass end of the black hole mass function. 
With future observational sensitivity in this domain, these models 
can be distinguished.

\begin{enumerate}

\item{Occupation fraction at $z = 0$: Our model for the formation of relatively high-mass black hole seeds
in high-$z$ halos has direct influence on the black hole occupation
fraction in galaxies at $z=0$. All our models predict that low surface brightness, bulge-less
  galaxies with large spin parameters (i.e. large discs) are systems
  where MBH formation is least probable.  We find that a significant fraction of low-mass
galaxies might not host a nuclear black hole. This is in very good
agreement with the shape of the $M_{\rm bh} - \sigma$ relation
determined recently from an observational census (an HST ACS survey)
of low mass galaxies in the Virgo cluster reported by Ferrarese et
al. (2006). While current data in the low mass
  regime are scant (Barth 2004; Greene \& Ho 2007; Kormendy \& Bender 2011), future instruments and surveys 
  are likely to probe this region of parameter space with significantly higher
  sensitivity.
}

\item{High mass end of the local SMBH mass function: While the models studied here (with 
different black hole seed formation
  efficiencies) are distinguishable at the low mass end of the BH mass
  function, at the high mass end the effect of initial seeds
  appears to be less important. These models cannot be easily distinguished 
  by observations at $z \sim 3$.}
  
  \end{enumerate}
  
  One of the key caveats of our picture is that it is unclear whether the
  differences produced by different seed models on observables at 
  $z = 0$ might be compensated or masked by BH fueling modes at earlier
  epochs. There could be other channels for BH growth that dominate at
  low redshifts like minor mergers, dynamical instabilities, accretion of
  molecular clouds and tidal disruption of stars. The decreased importance of the
  merger driven scenario is patent from observations of low-redshift AGN, 
  which are for the large majority hosted by undisturbed galaxies (e.g. Pierce 
  et al. 2007 and references therein) in 
  low-density environments. However, the 
  feasibility and efficiency of some alternative channels are still to 
  be proven, for example, the efficiency of feeding from large scale 
  instabilities (see discussion in King \& Pringle 2007; Shlosman, Frank \& Begelman 1989; 
  Goodman 2003; Collin 1999). 
  In any event, while these additional channels for BH {\it growth} can modify the 
  detailed shape of the mass function of MBHs, or of the luminosity function of 
  quasars, they will not create new MBHs.  The occupation fraction 
  of MBHs (see Figure 3) is therefore largely {\it independent} of the 
  accretion mechanism and a true signature of the formation process.
    
 To date, most theoretical models for the evolution of MBHs in galaxies
do not include {\it how} MBHs form. This work is a first analysis of
the observational signatures of massive black hole formation
mechanisms in the low redshift Universe, complementary to the
investigation by Sesana, Volonteri \& Haardt (2007), 
where the focus was on
detection of seeds at the very early times when they form, via
gravitational waves emitted during MBH mergers. We focus here on
possible dynamical signatures that forming massive black hole seeds
carry over to the local Universe. Obviously, the signatures of
seed formation mechanisms will be far more clear if considered jointly
with the evolution of the spheroids that they host. The mass, and
especially the frequency, of the forming MBH seeds is a necessary
input when investigating how the feedback from accretion onto MBHs
influences the host galaxy, and is generally introduced in numerical
models using extremely simplified, {\it ad hoc} prescriptions (e.g.,
Springel,  Di Matteo \& Hernquist 2005; 
Di Matteo, Springel \& Hernquist 2005; Hopkins et al. 2006;
Croton et al. 2005; Cattaneo et al 2006; Bower et al. 2006).
Adopting more detailed models for black hole seed formation, as outlined 
here, can in principle strongly affect such results. Incorporating  sensible 
assumptions for the masses and
frequency of MBH seeds in models of galaxy formation is necessary if
we want to understand the symbiotic growth of MBHs and their hosts.

\section*{Acknowledgements}

PN would like to acknowledge her collaborators Marta Volonteri and
Giuseppe Lodato with whom most of this work was 
done. She would also like to thank the John Simon Guggenheim Foundation
for support from a Guggenheim fellowship and the Institute for Theory
and Computation at Harvard University for hosting her.


\label{lastpage}
\end{document}